\documentclass[prl,aps,twocolumn,superscriptaddress,preprintnumbers,nopacs,floatfix,amsmath,amssymb]{revtex4}

\usepackage{graphicx}
\usepackage{dcolumn}
\usepackage{bm}

\usepackage{amsmath}
\usepackage{graphicx}
\usepackage{amssymb}
\usepackage{mathrsfs}
\usepackage{bbm}
\usepackage{epsfig}
\usepackage{makecell}

\newcommand{\be}{\begin{eqnarray}}
\newcommand{\ee}{\end{eqnarray}}

\begin{document}





%

%
\title{Solution X-ray scattering (S/WAXS) and 
structure formation in protein dynamics}

\author{Alexandr Nasedkin}
\email{alexandr.nasedkin@chalmers.se}
\affiliation{Department of Physics, Chalmers University of Technology, SE-412 96 Gothenburg, Sweden
}
\author{Jan Davidsson}
\email{jan.davidsson@kemi.uu.se}
\affiliation{Department of Chemistry, Uppsala University,
P.O. Box 803, S-75108, Uppsala, Sweden
}
\author{Antti J. Niemi$^{1,\!}$}
\email{Antti.Niemi@physics.uu.se}
\affiliation{Nordita, Stockholm University, Roslagstullsbacken 23, SE-106 91 Stockholm, Sweden}
\affiliation{Department of Physics and Astronomy, Uppsala University,
P.O. Box 803, S-75108, Uppsala, Sweden}
\affiliation{Laboratoire de Mathematiques et Physique Theorique
CNRS UMR 6083, F\'ed\'eration Denis Poisson, Universit\'e de Tours,
Parc de Grandmont, F37200, Tours, France}
\affiliation{School of Physics, Beijing Institute of Technology, Beijing 100081, P.R. China}
\affiliation{Laboratory of Physics of Living Matter, School of Biomedicine, Far Eastern Federal University, Vladivostok, Russia}
\homepage{http://www.folding-protein.org}
\author{Xubiao Peng}
\email{xubiaopeng@gmail.com}
\affiliation{Department of Physics and Astronomy, University of British Columbia, \\
Vancouver, British Columbia V6T1Z4, Canada}

\begin{abstract}
\noindent
We propose to develop mean field theory in combination with Glauber algorithm,
to model and interpret protein dynamics and structure formation in 
small to wide angle x-ray scattering (S/WAXS) experiments.  We develop the methodology by analysing 
the Engrailed homeodomain protein as an example. 
We demonstrate how to interpret S/WAXS data qualitatively, 
with a good precision, and over an extended temperature range.  We 
explain experimentally observed phenomena in terms of protein phase structure,  and we make 
predictions for future experiments how to address the data at different ambient temperature values. 
We conclude that a combination of mean field theory with Glauber algorithm 
has the potential to develop into a highly accurate, 
computationally effective and predictive tool for analysing S/WAXS data.  
Finally, we compare our results  with those obtained previously in an all-atom 
molecular dynamics simulation.
\end{abstract}



\maketitle

\section{Introduction}

The ability to reduce physical phenomena into simple fundamental laws does not
ensure the ability to reconstruct complex physical phenomena from these laws ~\cite{Anderson-1972}. 
An atomic level reconstruction  often encounters difficulties in large scale systems, and in particular when 
structural self-organisation takes place. 
In such scenarios various mean field theoretical descriptions, which  build 
on considerations of symmetry and its breaking 
in a physical system, can provide a pragmatic  
alternative ~\cite{Goldenfeld-1992}.  
Protein folding and dynamics is an example of a scenario, where the predictions of different  
approaches can be contrasted against each other. 
It is a setting where methods of mean field theory and the computationally highly demanding all-atom 
approach~\cite{Alm-1999,Baker-2000} can be compared, and possibly even employed conjointly.  

A mean field theory is most effective when the underlying physical system possess symmetries that have become broken.  
Indeed, 
various classification schemes ~\cite{Sillitoe-2013,Murzin-1995}
of crystallographic Protein Data Bank (PDB) structures reveal 
that folded proteins are built in an apparently symmetric, modular fashion. A symmetry principle based
mean field approach has been introduced, that identifies and models the individual building blocks of a protein 
in terms of soliton solutions of a 
generalised discrete nonlinear Schr\"odinger (DNLS) equation\cite{Chernodub-2010,Molkenthin-2011}.
The DNLS Hamiltonian is the paradigm integrable model  \cite{Faddeev-1987}, its soliton solution has found numerous physical
applications  \cite{kevrekidis}.   
Subsequently it has been proposed  that a combination of the ensuing generalised 
DNLS free energy with fluctuations accounted for using Glauber's description of 
nonequilibrium statistical mechanics ~\cite{Glauber-1963,Bortz-1975,Berg-2004} can provide both
an accurate and  a computationally effective 
way to model protein dynamics. At the same time, all-atom molecular dynamics (MD) simulations are reaching the maturity to  
describe the folding process of very fast-folding proteins ~\cite{Lindorff-Larsen-2011}. It has been shown
that an analysis of MD folding trajectory in terms of DNLS soliton and  concepts of mean field theory provides a
highly precise qualitative description  how protein folding and dynamics evolves  
during an all-atom simulation ~\cite{Dai-2016}.

Here we combine mean field theory including fluctuations computed by Glauber algorithm, 
to model protein dynamics in the context of small and wide angle 
solution X-ray scattering (S/WAXS) experiments. S/WAXS  is an experimental
technique that is emerging as a key tool to study proteins at low resolution  
~\cite{Doniach-2001,Putnam-2007,Rambo-2010,Blanchet-2013}.  
%
%
%
%
We propose that data obtained in a S/WAXS experiment 
can  be simulated and interpreted in a mean field based approach efficiently, 
with good precision, and over an extended temperature 
range.  For this we consider as an example the   {\it Drosophila melanogaster} 
Engrailed homeodomain (EnHD). Engrailed is a 61 residue fast-folding two-state  protein. It folds via a helical intermediate,
and forms a three helix bundle in the folded state.
Our reference structure has PDB code 2JWT ~\cite{Religa-2008}. This is a NMR structure, and we use the first entry in PDB
in our analysis. 

Previously the experimental S/WAXS results ~\cite{Nasedkin-2015}, on which we base our investigation, 
have been studied using all-atom molecular dynamics (MD) simulations ~\cite{Nasedkin-2015}. The  
Gromacs package ~\cite{Pronk-2013}
and Amber99SB-ILDN force field ~\cite{Lindorff-Larsen-2010} were employed, over an extended temperature range.  
%
%
%
%

We compare the results with mean field theoretical  
simulations in combination with finite temperature Glauber algorithm. 
We consider a
temperature range that covers and extends beyond the available experimental S/WAXS data.
We find that a mean field simulation yields  a very good description of the experimental observations,
and we make predictions for future experiments. A mean field theory based approach
significantly reduces the number  of degrees of freedom in comparison to an 
all-atom description, and as a consequence it proceeds several orders of magnitude faster:
The results presented here are obtained in a couple of hours {\it in silico} with a 
laptop computer.

\section{Methods}

\subsection{Mean field approach to proteins}
Our mean field  approach
is based on a Landau free energy ~\cite{Goldenfeld-1992} that relates to the C$\alpha$ backbone geometry. The time scale for 
a covalent bond oscillation is  around 10 ps, thus over biologically relevant
time scales the distance between two neighboring C$\alpha$ atoms can be approximated 
by the time averaged value 3.8 \AA. Accordingly, the skeletal C$\alpha$ bond $\kappa$ and torsion $\tau$ angles 
form a complete set of structural order parameters, to be employed  in the construction of a C$\alpha$ trace based 
Landau free energy \cite{Hinsen-2013}. The bond angles
are known to be relatively rigid and slowly varying, the differences $\Delta \kappa_i = \kappa_{i+1}-\kappa_i$ between
neighboring residues are
small. Thus the free energy $E(\kappa,\tau)$ can be expanded in the powers of the differences $\Delta \kappa_i $. 
A detailed analysis which builds on extensive symmetry considerations, in particular on the requirement that the 
functional form of the energy should remain invariant under local frame rotations,  shows ~\cite{Chernodub-2010,Molkenthin-2011,Niemi-2003,Danielsson-2010,Hu-2011a,Krokhotin-2012-1,Krokhotin-2012-2,Krokhotin-2013-2,Niemi-2014}
that in the limit of small variations  in $\Delta \kappa_i$  
the following expansion of the free energy can be used in the case of proteins
\[
 E(\kappa,\tau)  = 
\sum\limits_{i=1}^{N-1}  
\Delta \kappa_{i}^2  
+   \sum\limits_{i=1}^N \left\{ \lambda\, (\kappa_i^2 - m^2)^2  + \frac{d}{2} \,  \kappa_i^2  \tau_i^2  \right.
\]
\begin{equation}
\left. - b \kappa_i^2 \tau_i - a  \tau_i + \frac{c}{2} \tau_i^2 \right\}
\ + \sum\limits_{i\not=j}V(|\mathbf x_i - \mathbf x_j|)
\label{eq:A_energy}
\end{equation}
Here ($\lambda,m,a,b,c,d$) are parameters. 
For a given PDB protein structure the parameters are determined by training the 
minimum energy configuration of (\ref{eq:A_energy}) to model the PDB backbone. 

We recognise
in (\ref{eq:A_energy}) a deformation of the Hamiltonian that defines the discrete 
nonlinear Schr\"odinger (DNLS) equation ~\cite{Chernodub-2010,Molkenthin-2011}. The three first  terms coincide with a {\it naive} discretisation of the continuum nonlinear Schr\"odinger
equation. The fourth term ($b$) is the conserved momentum in the DNLS model,  the fifth ($a$) 
term  is the Chern-Simons term, and the sixth ($c$) term is the Proca mass; see
~\cite{Hu-2013,Ioannidou-2014,Ioannidou-2016,Gordeli-2016} for a detailed analysis of these contributions. Finally, the last term 
($V$) includes various long distance two-body interactions such as Coulomb and 
Lennard-Jones interaction between the residues. In the leading order this contribution
can be approximated
by a hard ball Pauli repulsion ~\cite{Chernodub-2010,Molkenthin-2011,Niemi-2003,Danielsson-2010,Hu-2011a,Krokhotin-2012-1,Krokhotin-2012-2,Krokhotin-2013-2,Niemi-2014}; see  ~\cite{Sinelnikova_2016} for more general long range interactions.

We validate (\ref{eq:A_energy}) qualitatively using Privalov's criteria:  According to ~\cite{privalov_1979,privalov_1989,Shaknovich_1989} 
the  folding of a protein should be cooperative, and it should resemble a first order phase transition. 
Indeed, the DNLS equation supports solitons, and solitons 
are the paradigm  cooperative organisers  in physical scenarios.
A soliton emerges as a solution to the  variational equations of (\ref{eq:A_energy}), and for this we first eliminate the torsion angles 
using the equation
\begin{equation}
\tau_i[\kappa] \ = \ \frac{a+b\kappa_i^2}{c+d\kappa_i^2}
\label{tau}
\end{equation} 
For bond angles we then  obtain
\begin{equation}
\kappa_{i+1} = 2\kappa_i - \kappa_{i-1} + \frac{dV[\kappa]}{d\kappa_i^2}  
 \label{kappaeq}
\end{equation}
where
\[
V[\kappa] = - \left( \frac{bc-ad}{d}\right) \, \frac{1}{c+d\kappa^2}  - \left( \frac{b^2 + 8 \lambda m^2}{2b}\right) \kappa^2 + \lambda \kappa^4
\]
The difference equation (\ref{kappaeq}) can be solved iteratively ~\cite{Molkenthin-2011}. The ensuing torsion angles are computed
from  (\ref{tau}), and the C$\alpha$ backbone coordinates are obtained by solving the discrete Frenet equation  \cite{Hinsen-2013,Hu-2011}.
A soliton solution models a super-secondary protein structure such as a helix-loop-helix motif,  and the loop corresponds to the soliton proper ~\cite{Niemi-2003,Chernodub-2010,Molkenthin-2011,Danielsson-2010,Hu-2011a,Krokhotin-2012-1,Krokhotin-2012-2,Krokhotin-2013-2,Niemi-2014}.

In order to reveal a relation between  (\ref{eq:A_energy}) and the structure of a first order phase transition, 
we note that in the case of a protein the  bond 
angles are rigid and the torsion angles are flexible. In particular,
the variations of $\kappa_i$ along the backbone are  small in comparison to changes in $\tau_i$: We may confirm from
(\ref{tau}) that a large change in values of $\tau_i$ entails a small change in values of $\kappa_i$ for parameters 
that are characteristic to protein backbones. Thus, over sufficiently large
distance scales we may try and proceed self-consistently, using only the mean values of the variables. For this   
we first solve for the  mean value of the bond angles $\kappa_i \sim \kappa$ in terms of the mean value of torsion angles 
$\tau_i \sim \tau$. From (\ref{eq:A_energy})  
\begin{equation}
\frac{\delta E}{\delta \kappa} = 0 \ \Rightarrow \ \kappa^2 \ = \ m^2 + \frac{b}{2\lambda} \tau - \frac{d}{4\lambda} \tau^2 \ 
\label{mftk}
\end{equation}
In those cases that are of interest to us, this equation always has a solution:  
Both $\kappa$ and $\tau$ are multivalued angular variables, 
and for proteins the parameters $b$ and $d$ are small in comparison with $m^2$ and $\lambda$.
We substitute the solution into (\ref{eq:A_energy}). For the energy this  gives
\begin{equation}
 -  \frac{d^2}{16 \lambda} \tau^4 + \frac{bd}{4\lambda} \tau^3 - \left( \! \frac{b^2}{\lambda} - 
2dm^2 - 2c \right) \tau^2 + \left( a+bm^2 \right) \tau 
\label{mftF}
\end{equation}
This is the canonical form of the Landau - De Gennes free energy for a first order phase transition, introduced originally
in the context of liquid crystals  ~\cite{DeGennes_1995}.
Thus  our qualitative validation of (\ref{eq:A_energy}) is complete, in the sense that we have confirmed that 
the free energy (\ref{eq:A_energy}) appears to be in line with the general arguments in  ~\cite{privalov_1979,privalov_1989,Shaknovich_1989}.

In Appendix A we elaborate on relations between the energy function (\ref{eq:A_energy}) and elastic network models.

\subsection{Solitons and Engrailed homeodomain} 

We  construct the C$\alpha$ trace of Engrailed homeodomain 
as a multi-soliton solution of (\ref{kappaeq}), (\ref{tau}). Our 
reference configuration is the first entry in the NMR structure with PDB code 2JWT.
There are seven individual solitons, including one at each of the flexible N and C terminals. The C$\alpha$ root-mean-square (RMS) 
distance  between the 2JTW and our multisoliton  is 0.67\AA, when 
we use  the parameter values that are given in Table \ref{table-1}. Note that there are in total 63 parameters including
the individual soliton centers, while there are 61 amino acids along the backbone: The presence of solitons enables us to combine 
the geometry of several amino acids into a single soliton profile, which greatly reduces the number of parameters in 
(\ref{eq:A_energy}). Since the number of parameters is {\it much} smaller
than the number of C$\alpha$ coordinates, the model has substantial predictive power that can be scrutinised in experimental scenarios.
\begin{table*}[h!]
\begin{center}
 \begin{tabular} {| p{1.2cm} | l | l | l | l | l | l | l | l  }
\hline
~ soliton & ~~ \ $\lambda_1$ & ~~ \ $\lambda_2$  & ~~ \ $m_1$  ~ & ~~ \ $m_2$  ~ \ & ~~~~ d  &  ~~~~ \ c  &~~~~ \ b  \\ \hline
~ \ 5-6   & ~ 2.888  ~  & ~ 1.454~   & ~ 1.174~ & ~ 1.462~  & ~ 1.061 e-09~  &  ~ 3.338 e-11~   & ~ 5.953 e-07~  \\ \hline
~ \ 8-9   & ~ 0.664   ~  & ~ 0.766~  & ~ 1.694~ & ~ 1.565~  & ~ 1.682 e-09~  &  ~ 1.667 e-09~   & ~ 2.574 e-07~  \\ \hline
~  10-11 & ~ 6.701 ~   & ~ 5.39~    & ~ 1.077~ & ~ 1.534~  & ~ 5.441 e-09~   & ~ 1.602  e-09~   & ~ 2.174 e-07~   \\ \hline
~ 23-24  & ~ 1.063 ~  & ~ 0.527~   & ~ 1.682~ & ~ 1.698~  & ~ 0.0~               &  ~ 1.234 e-08~   & ~ 1.721 e-07~   \\ \hline
~ 28-29  & ~ 7.737 ~  & ~ 6.699~   & ~ 0.849~ & ~ 1.463~  & ~ 6.187 e-09~   & ~ 1.438  e-09~   & ~ 6.766 e-08~   \\ \hline
~ 41-42  & ~ 0.37  ~   & ~ 0.839~   & ~ 1.674~ & ~ 1.566~  & ~ 0.0~                & ~ 1.009  e-09~   & ~ 3.21   e-08~   \\ \hline
~ 57-58  & ~ 7.09 ~    &  10.196~   & ~ 1.517~ & ~ 1.269~  & ~ 1.962 e-09~    & ~ 9.333  e-09~   & ~-2.377  e-07~   \\ \hline
\end{tabular}
\end{center}
\caption{The  parameters in the energy function (\ref{eq:A_energy}) for 2JWT. The first column defines the center of each soliton, in terms of residue number. For the
parameter  $a$ we use the fixed value 1.0 e-07, which determines the relative scale between the bond and torsion angle flexibility.}
\label{table-1}
\end{table*}

\begin{figure}[h]
  \vspace{0.2cm}  \resizebox{8 cm}{!}{\includegraphics[]{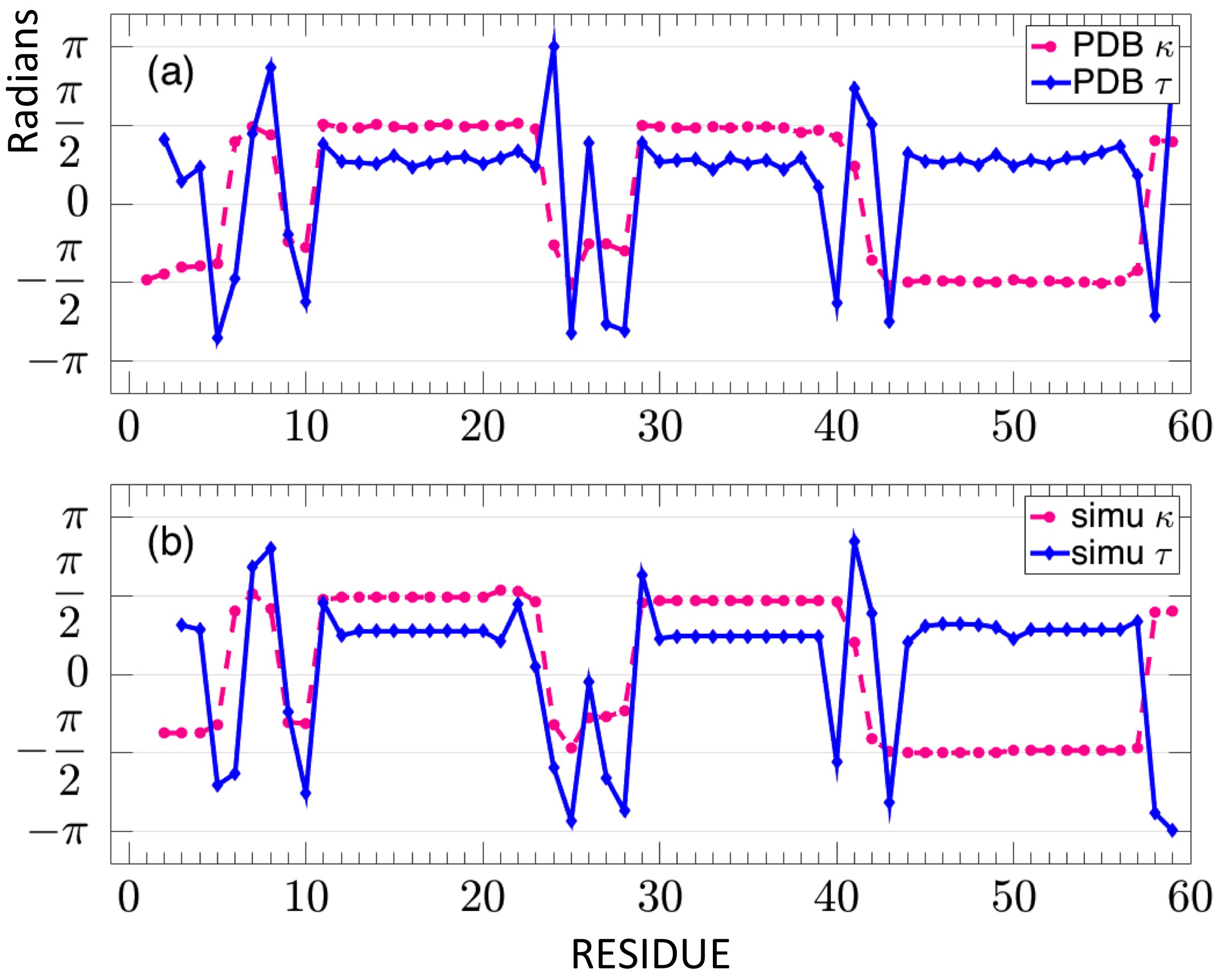}}
\caption {\small  Color online) (Top) Comparison of bond ($\kappa_i$)  angles in the first NMR entry of the PDB structure 2JWT and in the ensuing multisoliton solution.  (Bottom) Comparison of torsion ($\tau_i$)  angles in the first NMR entry of the PDB structure 2JWT and in the ensuing multisoliton solution; note that the torsion angle is defined $\mod(2\pi)$ thus the apparent differences at sites $i=24$ and $i=59$ are smaller than what  they appear.  
      }   
          \label{fig-1}
 \end{figure}

\begin{figure}[h]
  \vspace{0.2cm}  \resizebox{8 cm}{!}{\includegraphics[]{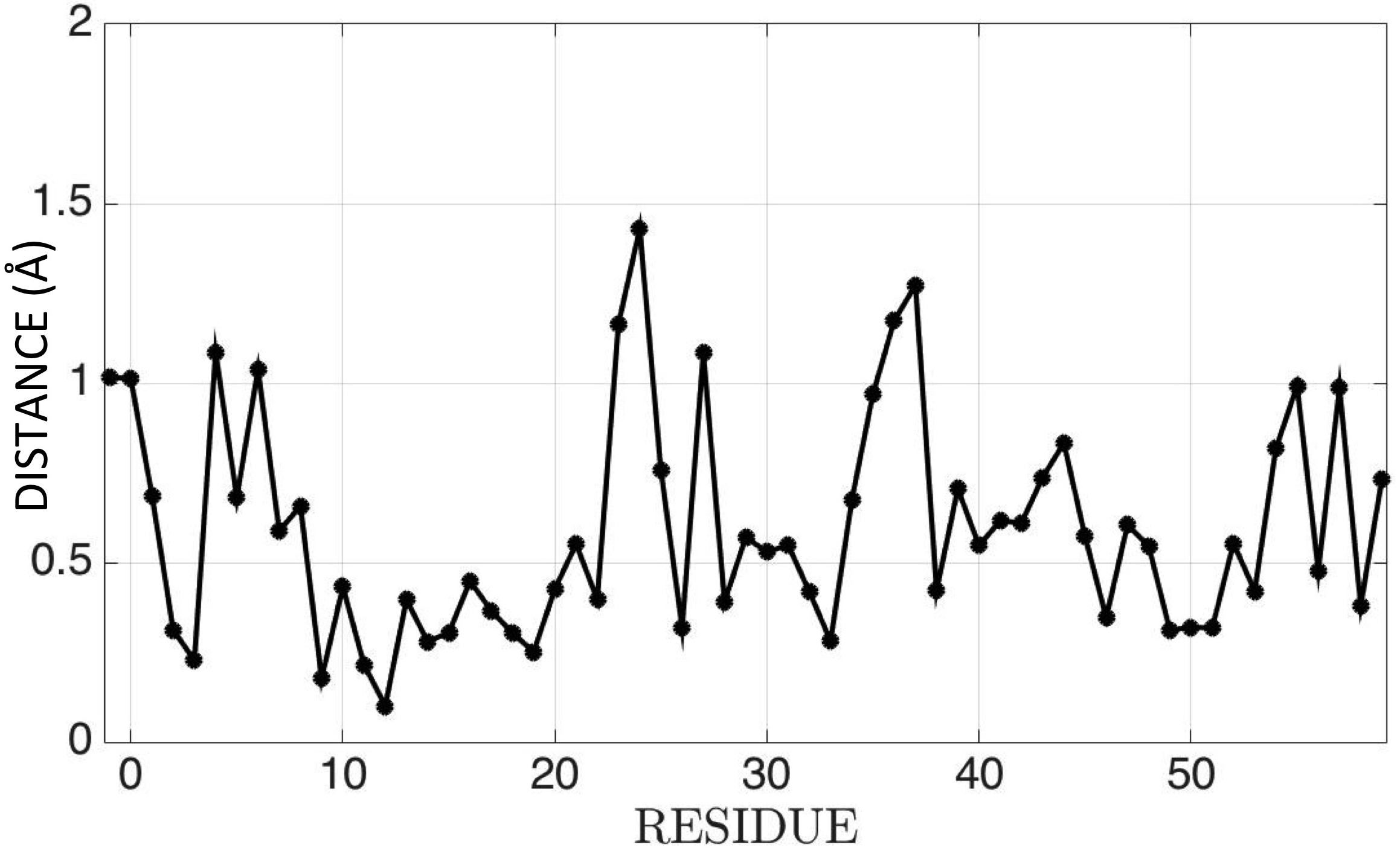}}
\caption {\small  Color online) The residue-wise distance between the C$\alpha$ atoms in the PDB structure 2JWT (first entry) and the
      ensuing multisoliton; the grey area corresponds to a 0.2 \AA~ zero point fluctuation distance.
      }   
                 \label{fig-2}
 \end{figure}

\begin{figure}[h]
  \vspace{0.2cm}  \resizebox{7 cm}{!}{\includegraphics[]{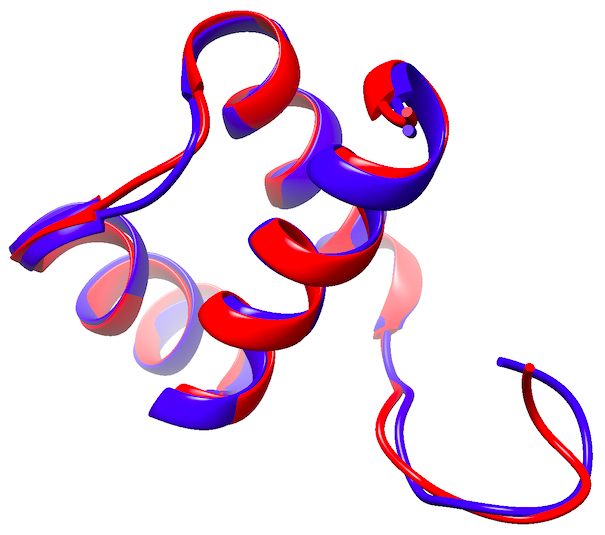}}
\caption {\small   (Color online) The interlaced C$\alpha$ traces of the PDB structure 2JWT (first entry) (in red) and the ensuing multisoliton (in blue).
      }   
                 \label{fig-3}
 \end{figure}

In Figure \ref{fig-1} we compare the bond and torsion angle spectra in the PDB structure and the multisoliton.
%
%
%
%
%
%
%
%
%
%
%
%
%
%
%
%
%
%
%
%
%
%
%
%
In Figure \ref{fig-2} we show the distance between the individual C$\alpha$ atoms in the PDB structure and the multisoliton. 
%
%
%
%
%
%
%
%
%
%
%
%
%
%
%
%
%
%
%
%
%
%
%
%
In Figure \ref{fig-3} we show the three dimensional interlaced C$\alpha$ traces for the PDB structure and the multisoliton.
%
%
%
%
%
%
%
%
%
%
%
%
%
%
%
%
%
%
%
%
%
%
%

\subsection{Soliton and Glauber algorithm}

We study how the present soliton model of Engrailed  responds to variations in ambient temperature using 
the Glauber algorithm ~\cite{Glauber-1963,Bortz-1975,Berg-2004}. We justify the Glauber algorithm with the following
line of arguments: In the case of a simple spin chain the Glauber algorithm reproduces  Arrhenius law, and
the folding of many  short protein chains follows  Arrhenius law ~\cite{Alm-1999}.  Since Engrailed is a relatively short 
two-state fast-folding protein ~\cite{Mayor-2003}, its folding should obey Glauber dynamics with good accuracy.

Indeed, Glauber algorithm has a claim to universality in the sense that  Glauber  dynamics 
approaches Gibbsian equilibrium distribution at an exponential rate, as expected in a near equilibrium system
 ~\cite{Glauber-1963,Bortz-1975,Berg-2004}.

We perform nine simulations at the Glauber temperature factor values 
$k\theta = 10^{-12}, 10^{-11},  \dots, 10^{-4}$. Each simulation involves
$10^6$ Monte Carlo  steps and every 1.000$^{th}$ structure is chosen 
for sampling.  We have carefully tested the algorithm length to ensure that we model full thermalisation and
for simulation details we refer to ~\cite{Krokhotin-2013-2,Krokhotin-2013-1}.
Since the Landau free energy only engages the C$\alpha$ atoms, we employ
the {\it Pulchra} reconstruction program ~\cite{Rotkiewicz-2008} to generate the 
all-atoms structures.

\begin{figure}[h]
  \vspace{0.2cm}  \resizebox{8 cm}{!}{\includegraphics[]{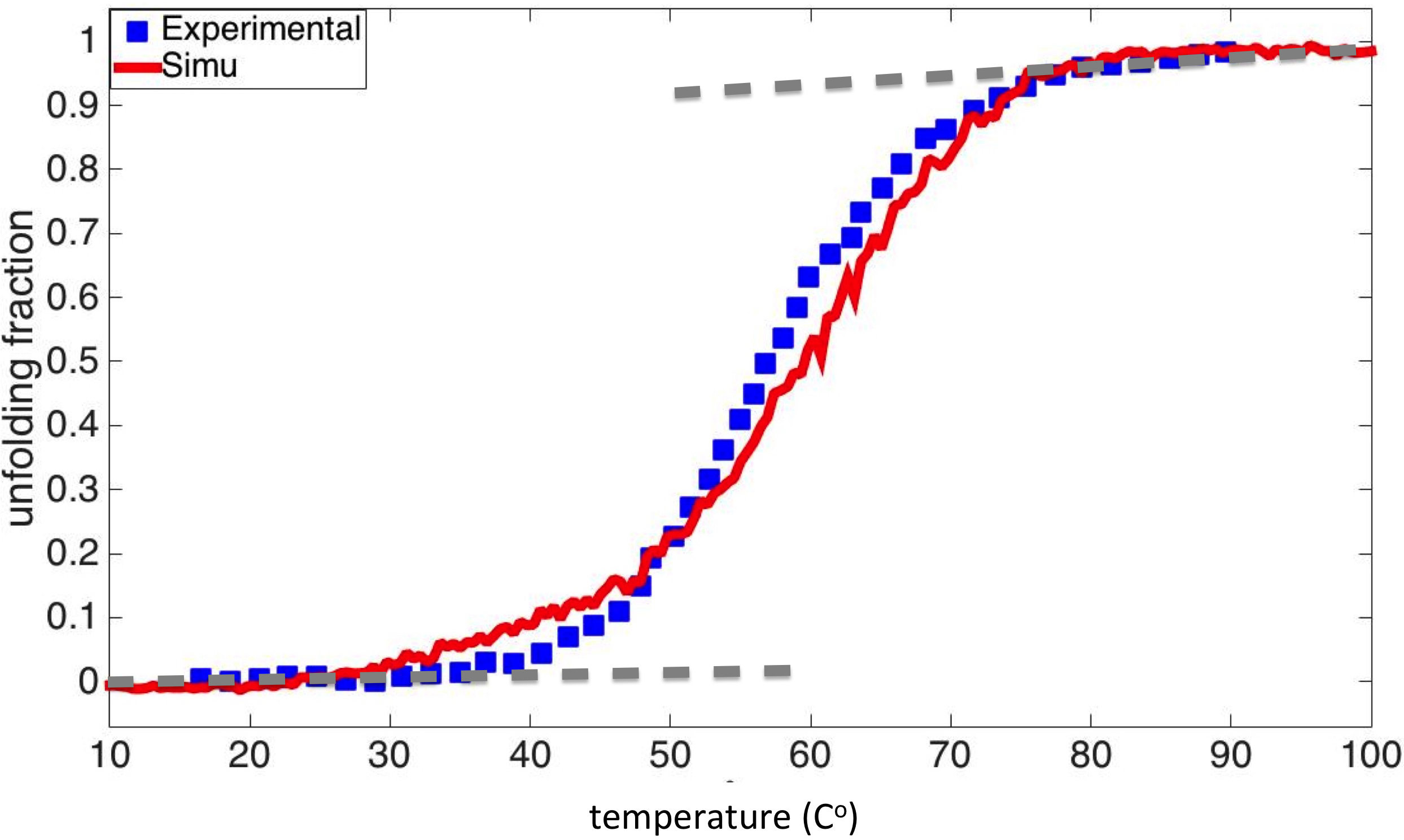}}
\caption {\small   (Color online) Comparison of simulated temperature dependence in 
      $\alpha$-helical content with temperature dependence of unfolded fraction observed using 
      222 nm CD spectra in wild type Engrailed. The experimental  data is adapted
      from Figure 3 (bottom) of ~\cite{Ades-1994}. 
      }   
        \label{fig-4}
 \end{figure}

The Glauber temperature factor $k\theta$ does not coincide with the physical temperature $T$ (measured in Kelvin). 
However,  the two
can be related by a renormalisation procedure. General arguments presented in ~\cite{Krokhotin-2013-1} suggest that the 
relation between the  two should have the form 
\begin{equation}
k\theta = T^\gamma e^{\alpha T - \beta}
\label{ren-T}
\end{equation}
Unfortunately,  the S/WAXS data that is available to us in the case of Engrailed, is not sufficient to determine the parameters
in (\ref{ren-T}).  More data is needed, over a more extended temperature range.  We hope that future experiments can 
provide us with such data.  Meanwhile, we  rely on other experimental techniques to deduce the parameters
$\alpha$, $\beta$ and $\gamma$ in (\ref{ren-T}):

Circular dichroism (CD) spectroscopy  can 
measure the  helical content of a protein, as a function of temperature~\cite{Kelly-2005}. Accordingly,
we can determine the parameters in (\ref{ren-T}) by comparing the temperature dependence between experimentally 
observed and simulated helical content. 
For experimental data we  use the CD results on Engrailed that are reported in Figure 3 (bottom) of ~\cite{Ades-1994}.
For simulation, we first deduce from a statistical analysis of PDB structures that a 
C$\alpha$ atom which is centered at $\mathbf r_{i}$ can be taken to be in an $\alpha$-helical position when
\[
| \mathbf r_{i+4} - \mathbf r_{i} |\approx  6.2 \ \pm 0.5 \  {\rm \AA} \ \ \ \ \& \ \ \ \  |\tau_i - \tau_{0}| < 0.6 \  {\rm (rad)}
\] 
where $\tau_{0}$ is the PDB average value of the $\alpha$-helical torsion angle. 
In  Figure \ref{fig-4} we compare  the data that we infer from ~\cite{Ades-1994}   with  
$k\theta$  dependence of $\alpha$-helical content in our Glauber algorithm simulations.
%
%
%
%
%
%
%
%
%
%
%
%
%
%
%
%
%
%
%
%
%
%
%
From the Figure  (\ref{fig-4}) we deduce the following  relation (\ref{ren-T}),
\begin{equation}
k\theta \ = \ T^{\, 0.92} e^{ 0.113 {\times} T -60.9}
\label{ren-T2}
\end{equation}
where the physical (experimental) temperature $T$ is measured in Kelvin. For convenience, the conversion
between those Glauber temperature factor values that we use in our simulations 
and Celsius degrees are shown in Table \ref{table-1}

In the Figure \ref{fig-4} we also display our estimates for the high temperature and low temperature linear asymptotes. 
We estimate that there is an onset of linear behaviour above the high ($T_h$)  and below low ($T_l$)  temperature values
(order of magnitude)
\begin{equation}
\begin{matrix}
T_h \ \approx \ 75 \ ^\mathrm o \mathrm C \\
T_l \ \approx \ 30 \ ^\mathrm o \mathrm C
\end{matrix}
\label{lh}
\end{equation}
%
%
%
%
%
%
%
{
\begin{table*}[htb]
       \centering
\begin{tabular}{|c||c|c|c|c|c|c|c|c|c|}
\hline  $k\theta$ & 1.0\,e-12 & 1.0\,e-11 & 1.0\,e-10 & 1.0\,e-9 & 1.0\,e-8 & 1.0\,e-7 & 1.0\,e-6 & 1.0\,e-5 & 1.0\,e-4   \tabularnewline
\hline 
                $^o$C        &      $\sim$ -25   &   $\sim$ -5    &      $\sim$ 15        &      $\sim$ 35     &      $\sim$ 55      &     $\sim$ 75      
                &     $\sim$ 95       &     $\sim$ 115      &    $\sim$ 135    
\tabularnewline
\hline \end{tabular}
\caption{\small
\it Conversion between the simulated Glauber temperature factor  values $k\theta$ 
and the physical temperature values measured in Celsius, 
obtained using (\ref{ren-T2}).
 }
       \label{table-2}
\end{table*}
}
%
%
%
%
%
%
%
%
%
%

\subsection{Computations of S/WAXS profiles}
We analyse the results from experimental scattering data computations presented  in ~\cite{Nasedkin-2015}. These computations
were performed with CRYSOL software which is 
part of the ATSAS package ~\cite{Petoukhov-2012}. Scattering spectra were calculated in the range of collected 
S/WAXS spectra, with the solvent density  \textit{dns} set to $340 e/nm^3$.
The number of spherical harmonics that define the resolution of the scattering curve \textit{lm} was set to 
seven. All the other settings of CRYSOL were set to the default values.
Scattering spectra were calculated from protein structures containing all the atoms. For details 
see ~\cite{Nasedkin-2015}. 

\subsection{Experimental S/WAXS}
The scattering patterns that we use were collected at the cSAXS beamline of the Swiss Light 
Source (SLS) facility at PSI, Switzerland. The EnHD was expressed and purified as previously described in ~\cite{Ades-1994}. A moderate 
concentration 
of 1.1mM EnHD was dissolved in buffer containing 50 mM HEPES and 100 mM NaCl at ph=8.0. Protein sample was constantly pumped 
during the 
experiment allowing fresh sample exposed to the X-rays at all time and thus reducing risk of high-dosage agglomerations. Collected 
scattering spectra in 
the range between $q_{min}=0.07$~\AA~and $q_{max}=0.71$~\AA~were used for further analysis; here $q=4\sin(\theta \ / \lambda)$ is the 
scattering 
vector, with $2\theta$ the scattering angle of incoming X-rays and $\lambda$ is the X-ray wavelength.

Scattering pattern at every experimental temperature in ~\cite{Nasedkin-2015} 
is a sample average of separate spectra each of them accumulated during 10 seconds. 
Datasets were filtered by outlier rejection and tested for possible agglomerations. The experimental error of a scattering curve was calculated as a standard deviation across each dataset.

\subsection{Optimization algorithm}
We  fit the simulated  S/WAXS spectra to the experimental data in ~\cite{Nasedkin-2015}. For this we
use the ensemble optimization method ~\cite{Bernado-2007},  it identifies the ensemble of structures which are 
best representing the experimental spectrum. Fitting is scored based on its $\chi^2$ value indicating a 
difference between the experimental and theoretical scattering profiles; 
see for example ~\cite{Schneidman-Duhovny-2010}.

We fit each of the theoretically simulated pool to the scattering profiles at every available 
experimental temperature. We perform optimization runs with 200 iterations, and every iteration returns
10 best-fitted ensembles that we use at the next step. Each ensemble contains a maximum of 
20 spectra. We tune mutation and crossing operators  for the fastest convergence of 
the fitting ~\cite{Nasedkin-2015}. The code for the optimization algorithm is implemented in a 
MATLAB package ~\cite{package}.

\subsection{Comparative all-atom molecular dynamics (MD) simulations}
We compare our mean field simulation results with  a pool of structures that were obtained in an earlier 
all-atom molecular dynamics investigation ~\cite{Nasedkin-2015}. Those  MD simulations were performed using the GROMACS
package ~\cite{Pronk-2013},  and the AMBER99SB-ILDN force field  ~\cite{Lindorff-Larsen-2010}  with the TIP3P water model ~\cite{Jorgensen-1983};
the protein structure was kept at normal pressure, with constant temperature values. 
Eight MD trajectories were generated for conformational sampling, each trajectory had a length of 100 ns 
and sampling was performed every 10 ps. This resulted 
in 10.000 sampled protein conformations at simulation 
temperature values $t = 275, \ 300, \ 325, \ 350, \ 375, \ 400, \ 450, \ 500\  \mathrm K$
{\it i.e.} $t \approx 0, 25, 50, 75, 100, 125, 175$ and $225$ $^\mathrm o$C. Note
that these structures were generated {\it solely} for the purpose of sampling the conformational landscape.
Thus the fact that very high and unphysical temperature values were used  in some of the simulations is not an issue.  
In order to compare the simulation results with 
S/WAXS data, the large number of MD conformers was reduced by clustering them,  using
structural similarity as a criterion.  The central cluster structures were then used as a
representative of each cluster, and employed in ensemble fitting with experimental data ~\cite{Nasedkin-2015}.

\section{Results}

\noindent
We perform comparisons between the mean field model simulations and experimental data, 
using a number of different criteria:

\subsection{Radius of gyration}

The  literature ~\cite{DeGennes_1979,Grosberg_1994,Schafer_1999} commonly attributes
three different phases to linearly conjugated  
polymers; see however ~\cite{Sinelnikova_2016}:
At low temperatures a protein structure is expected to reside in the space filling collapsed 
phase where attractive forces dominate. At high temperatures where repulsive interactions
prevail, a protein structure is in the self-avoiding random walk phase. Between these two
phases there is a transition region where the attracting and repelling interactions balance each other, and
the protein structure should resemble an ordinary random walk.

\begin{figure}[h]
  \vspace{0.2cm}  \resizebox{8 cm}{!}{\includegraphics[]{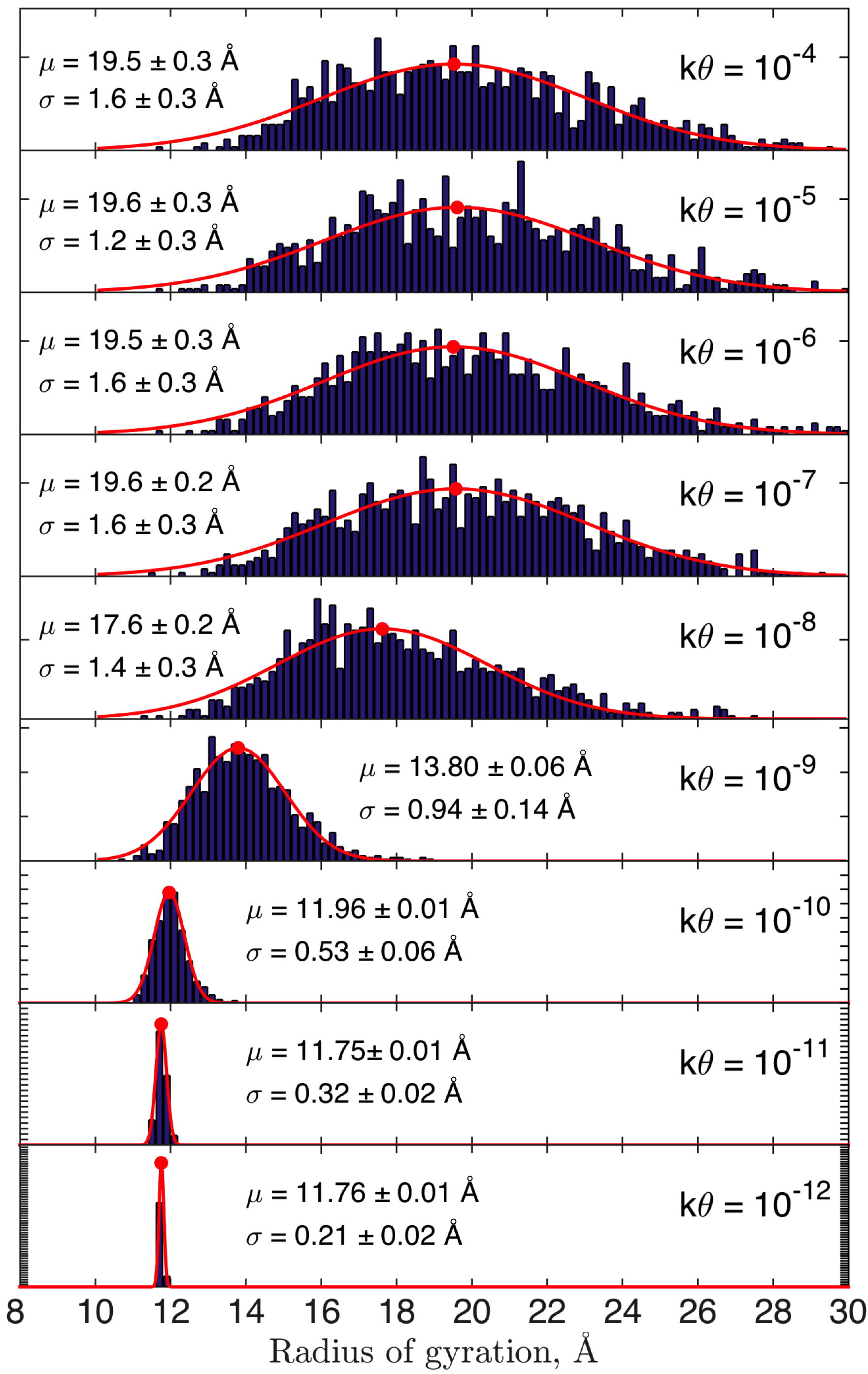}}
\caption {\small  (Color online) Distribution of the radii of gyration in mean field  model simulations at different Glauber temperature factor values.
      }   
                \label{fig-5}
 \end{figure}

\begin{figure}[h]
  \vspace{0.2cm}  \resizebox{8 cm}{!}{\includegraphics[]{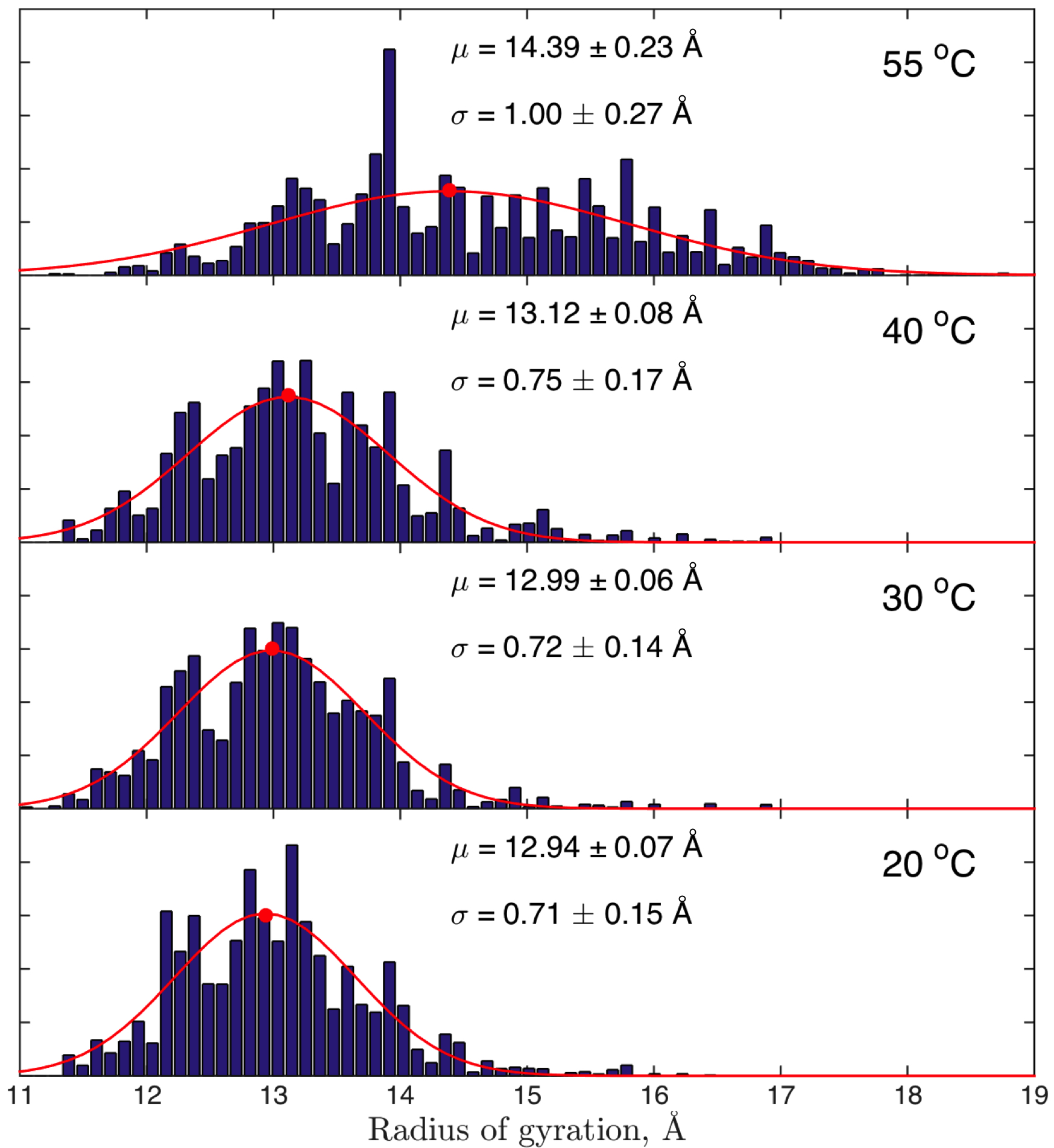}}
\caption {\small  (Color online) Distribution of the radii of gyration, as recovered from the experimental S/WAXS  data at different temperatures.
      }   
                      \label{fig-6}
 \end{figure}

The radius of gyration $R_g$ is a widely used order parameter, to determine the phase structure ~\cite{DeGennes_1979,Grosberg_1994,Schafer_1999}.
In Figure \ref{fig-5}  we show the mean field model
%
%
%
%
%
%
%
%
%
%
%
%
%
%
%
%
%
%
%
%
%
%
%
distribution of $R_g$, for each of the nine Glauber temperature  factor values that we use in our simulations.
In Figure \ref{fig-6} we show the experimentally measured  S/WAXS values of $R_g$, at different temperatures. 
%
%
%
%
%
%
%
%
%
%
%
%
%
%
%
%
%
%
%
%
%
%
%
We note that both in the case of the mean field model and the S/WAXS experiments the
$R_g$ distribution  has a shape which is reminiscent of a single Gaussian. 
Moreover, in the case of the mean field model we identify three different regions in Figure \ref{fig-5}:    

\vskip 0.1cm
\noindent
$\bullet$ For the very small temperature  factor values $k\theta = 10^{-12}$ - $10^{-10}$ corresponding to physical temperatures below $\sim 15 \, ^\mathrm o\mathrm C$,  
the mean value of radius of gyration
is stable with value around
$<\!\!R_g\!\!> \approx 11.8 - 11.9 \ $\AA. The variance $\sigma$ is small and increases from 
$\sigma \sim 0.05\ $\AA ~to $\sigma \sim 0.3\ $\AA~ with increasing $k\theta$.

\vskip 0.1cm
\noindent
$\bullet$ The intermediate temperature factor values between $10^{-9}$ and
$10^{-8}$ covers the physiologically interesting temperature range $\sim 35-55 \, ^\mathrm o\mathrm C$  and constitutes 
a transition region where the mean value of $R_g$ increases first to $<\!\!R_g\!\!> \approx 13.8\ $\AA~ at  
$k\theta = 10^{-9} $ 
and then to $<\!\!R_g\!\!> \approx 17.6\ $\AA~ at $k\theta = 10^{-8}$. The variance increases similarly, 
first to $\sigma \approx 0.9\ $\AA~ and then to $\sigma \approx  1.9 \ $\AA. 

\vskip 0.1cm
\noindent
$\bullet$ Finally, when $k\theta = 10^{-7}$ and above corresponding to physical temperature values above $\sim 75  \, ^\mathrm o\mathrm C$, 
the distribution becomes stabilised with mean value $<\!\!R_g\!\!> \approx 19.6\ $\AA~ as the variance converges 
towards the high temperature limit $\sigma \approx  2.5\ $\AA. 

\vskip 0.1cm
Apparently, the  temperature factor values $k\theta = 10^{-12} -  10^{-10}$
correspond to the low temperature collapsed phase. The  values $k\theta = 10^{-7} \ \sim 75 ^\mathrm o \mathrm C$ 
and above correspond to the high 
temperature self-avoiding random walk phase.  Values in the range $10^{-9} - 10^{-8}$ $\sim 35-55 \, ^\mathrm o\mathrm C$ are in 
the intermediate random walk transition region. See also the experimentally measured CD spectrum  ~\cite{Ades-1994}, 
shown in Figure \ref{fig-4}.

\vskip 0.2cm
\noindent
In the experimental S/WAXS data shown in Figure \ref{fig-6} we identify two different regimes: 

\vskip 0.1cm
\noindent
$\bullet$ When T is in the range
$20 \, ^{\rm o}\mathrm C \ - \ 30 \, ^{\rm o} \mathrm C$ the value of $R_g$ has a very small temperature dependence,  with mean
$<\!\! R_g\!\! > \approx 12.9-13.0 \ $\AA ~ and variance that is similarly essentially temperature independent
with $\sigma \approx 0.5 \ $\AA.  

\vskip 0.1cm
\noindent
$\bullet$ Between $40 \, ^{\rm o} \mathrm C \ - \ 55 \, ^{\rm o} \mathrm C$ there is an onset of a  
transition: Both $R_g$ and $\sigma$ start increasing so that when  T~=~55 $^{\rm o}$C we have 
$<\!\! R_g\!\! > \approx 14.4\ $\AA ~ and $\sigma \approx 1.0\ $\AA.

\begin{figure}[h]
  \vspace{0.2cm}  \resizebox{8 cm}{!}{\includegraphics[]{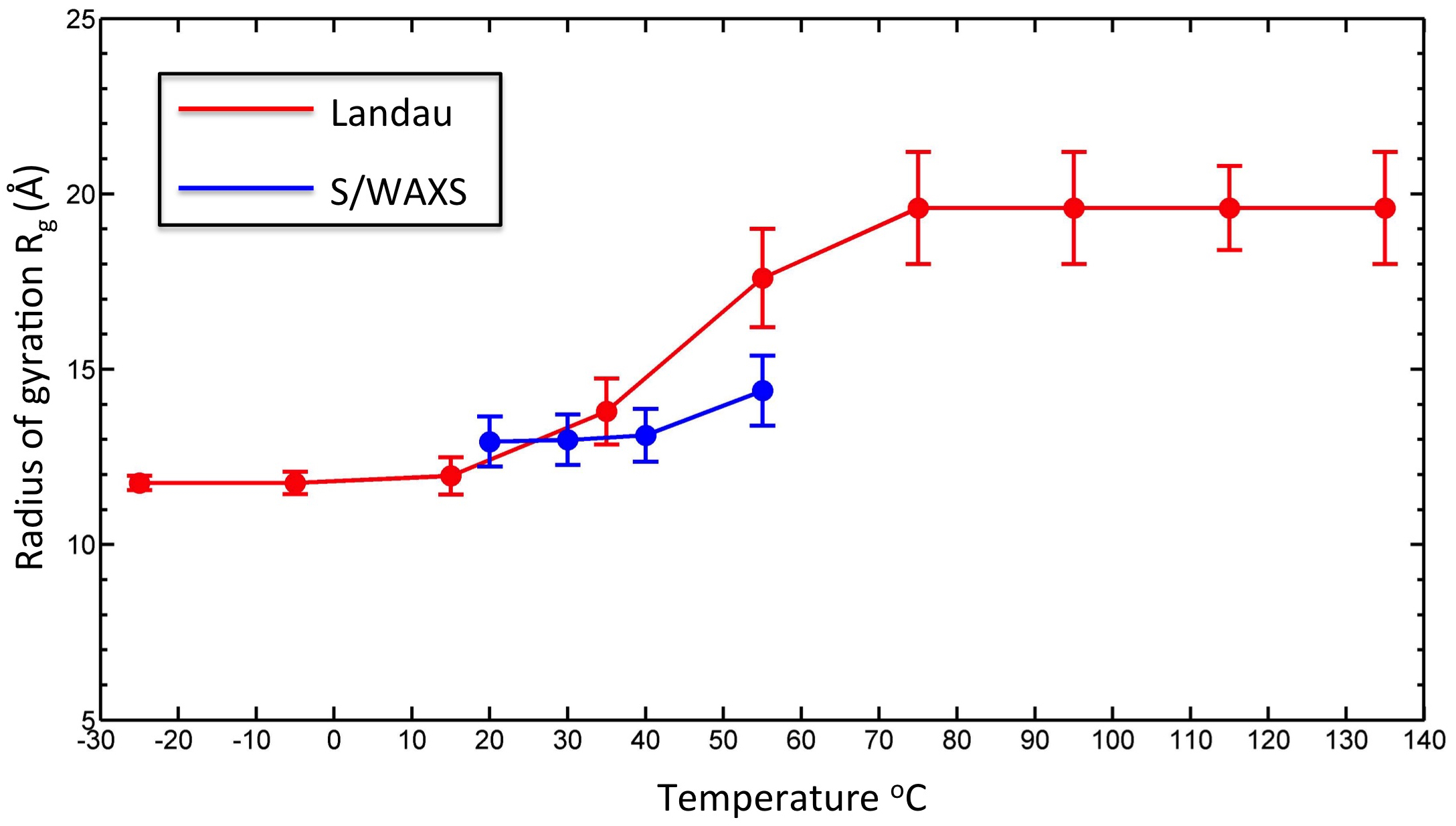}}
\caption {\small   (Color online) Comparison between the temperature dependence of radius of gyration in the mean field model and in the S/WASX 
      data. The error-bars denote the one $\sigma$ of the Gaussian fits, in Figures \ref{fig-5} and \ref{fig-6} respectively. 
      }   
                             \label{fig-7}
 \end{figure}

\vskip 0.2cm
\noindent
Figure \ref{fig-7}  shows a comparison 
between the mean field model results and the experimental S/WAXS data.
%
%
%
%
%
%
%
%
%
%
%
%
%
%
%
%
%
%
%
%
%
%
%
%
%
%
%
%
%
%
%
We observe that the T~=~$55 \, ^{\rm o}\mathrm C$ experimental S/WAXS structures start approaching
the self-avoiding random walk phase.  Apparently,  
the fully developed self-avoiding random walk phase is not visible in the temperature range which is covered by the 
S/WAXS data available to us, according to mean field
model.  See also Figure \ref{fig-4}.

\vskip 0.2cm
\noindent
Finally, in Figure \ref{fig-8}  we show the $R_g$ values from the MD simulation, adapted from ~\cite{Nasedkin-2015}. We observe the following:
%
%
%
%
%
%
%
%
%
%
%
%
%
%
%
%
%
%
 \begin{figure}[h]
  \vspace{0.2cm}  \resizebox{7.5cm}{!}{\includegraphics[]{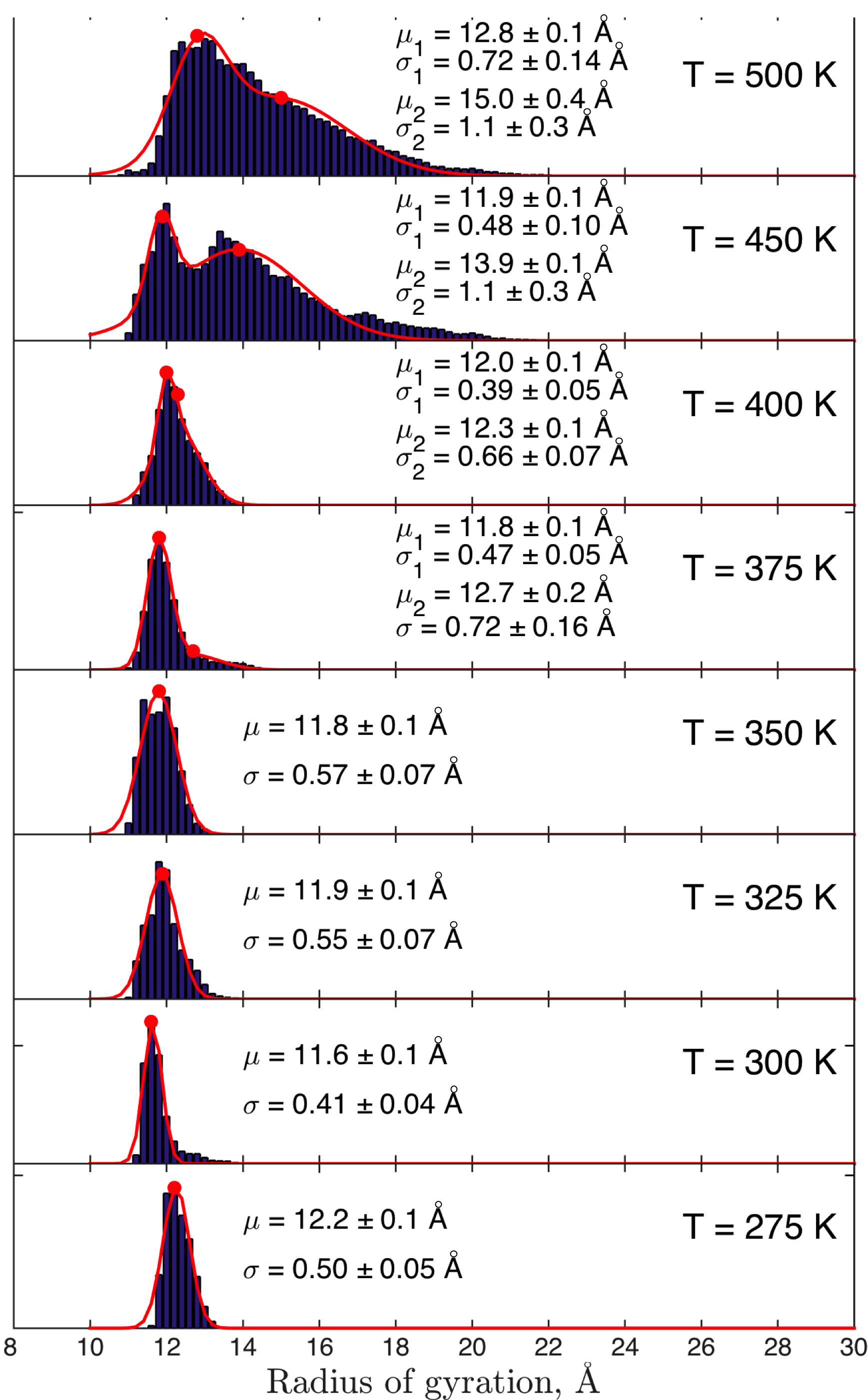}}
\caption {\small  (Color online) Distribution of the radii of gyration in the MD simulations.
 }   
                             \label{fig-8}
 \end{figure}
%
%
%
%
%
%
%
%
%
%
%
%

\vskip 0.1cm
\noindent
$\bullet$ The results up to $T \sim 400$K $\sim 125 \ ^\mathrm o \mathrm C$ are  very similar
to the collapsed phase 
low temperature mean field model results. In particular the  $T=325$K and $T=350$K results
are very similar to the $k\theta = 10^{-10} $ mean field model results, at the level  of the Gaussian distributions.  

\vskip 0.1cm
\noindent
$\bullet$ Between $T=400$K and $T=450$K there is an apparent transition, and both in the 
$T=450$K and in the $T=500$K data profiles
we identify a mixture of two Gaussian distributions:  
There is one Gaussian peaked at  $T=450$K which is akin
the one at collapsed phase in the mean field model. There is one Gaussian peaked at  $T=500$K which is akin the experimentally 
observed
low temperature T~=~290K and 310K distributions 
in the mean field model.
There is second Gaussian peaked at  $T=450$K which resembles the $k\theta =$  310K mean field model distribution. There  
is also a second  Gaussian peaked at  $T=500$K which is quite close to the T~=~55 $^{\rm o}$C experimental distribution.

\vskip 0.1cm
We propose that the double Gaussian distributions that we identify at $T=450$K and $T=500$K, 
reflect the {\it enormous} computational complexity of MD simulations: The initial configuration that is used 
in the MD
simulations is the experimental reference structure with PDB code 2JWT. Apparently, the available 
computational resources are not quite sufficient to observe the development of a fully thermalised single Gaussian 
distribution, akin those we find in the mean field model simulations and the S/WAXS experiment. 

We note that the transition in MD simulations between $T=400$K and $T=450$K is 
in line with ~\cite{Lindorff-Larsen-2011} where  the value $T_c = 390 \pm 7 $K is reported for the 
melting temperature of Engrailed homeodomain protein.

\subsection{Goodness-of-fit analysis}

%
%
%
%
%

The radius of gyration analysis which is summarised by Figure \ref{fig-7} proposes that the available experimental S/WAXS data 
in ~\cite{Nasedkin-2015} is best described by the mean field model when the Glauber temperature factor values 
are in the range $10^{-10}  - 10^{-9} $ {\it i.e.} $\sim 15 - 35 \ ^\mathrm o \mathrm C$. In particular, 
the physiologically relevant  $k\theta =  10^{-9} \sim 35 \ ^\mathrm o \mathrm C$ simulation appears to have a good fit
to experimental data, at this temperature.

\begin{figure}[h]
  \vspace{0.2cm}  \resizebox{8 cm}{!}{\includegraphics[]{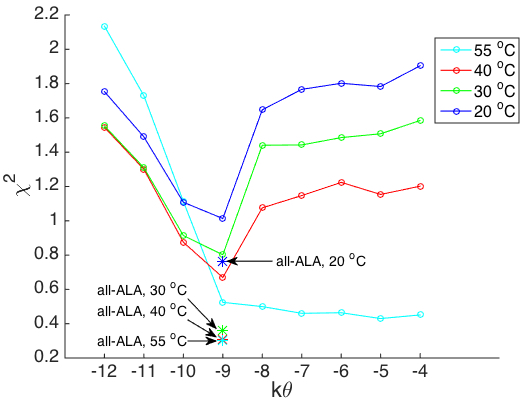}}
\caption {\small   (Color online) Dependence of the $\chi^2$ values obtained by fitting mean field model simulations, complemented 
      with {\it Pulchra} all-atom reconstruction.  The solid lines connecting circles are a guide for the eye. Also shown separately are $\chi^2$ values we obtain using a simplified C$\alpha$-C$\beta$ (mock alanine) structure in {\it Pulchra}.
      }   
                             \label{fig-9}
 \end{figure}

In Figure \ref{fig-9} we present a  {\it goodness-of-fit} ($\chi^2$) analysis. The 
Figure displays the  $\chi^2$ values we obtain in the mean field model, when we use {\it Pulchra}
~\cite{Rotkiewicz-2008}
to complete the C$\alpha$ backbone into an all-atom model. The result is shown independently  for each of the four 
available experimental  S/WAXS temperature value, as a function of the  temperature factor $k\theta$.
%
%
%
%
%
%
%
%
%
%
%
%
%
%
%
%
%
%
%
%
%
%
%
%
%
%
%
%
%
%
%
%
%
%
%
%
%
%
%
%
%
%
%
%
%
%
%
%
%
%
%
%
%
%
 \begin{figure}[h!]
  \vspace{0.2cm}  \resizebox{7.5cm}{!}{\includegraphics[]{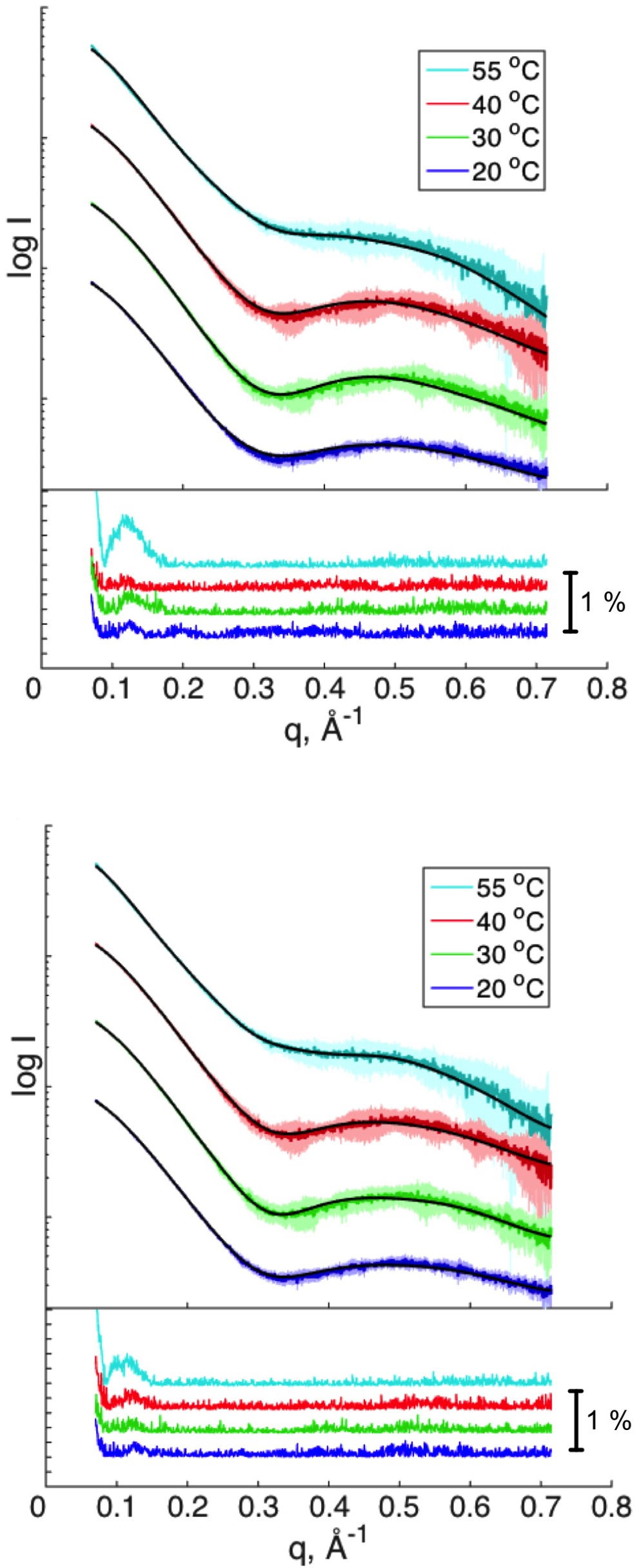}}
\caption {\small   (Color online)  (Top) Fitting of the mean field model pools generated at $k\theta = 10^{-9}$ with the experimental S/WAXS. 
      (Bottom) Same as top,  for the mixed temperature MD pool with the experimental S/WAXS. Plots are shifted vertically to 
      enhance visibility.  
      The experimental errors are shown in pale colour. The inset (second plot) in top and bottom Figures
      shows the difference between experimental and fitted profiles. The inset corresponds to one percent of the scattering intensity 
      extrapolated at $q$=0.     
      }   
                             \label{fig-10}
 \end{figure}
%
%
%
%
%
%
%
%
%
%
%
%
%

The results from the $\chi^2$ analysis shown in Figure \ref{fig-9} are in line with 
the conclusions from the radius of gyration analysis: 
In the case of mean field  model with all-atom {\it Pulchra} reconstruction, 
for the $20 \, ^{\rm o}\mathrm C \, - \, 40\, ^{\rm o}\mathrm C$ experimental
S/WAXS data the minimum  value of
$\chi^2$ occurs in the vicinity of the mean field model temperature factor value 
$k\theta = 10^{-9}$. The corresponding $\chi^2$ values are low, in the range $\sim 0.7 - 1.0$.  
For the highest temperature T=55 $^{\rm o}$C experimental S/WAXS data,  the $k\theta = 10^{-9}$ mean field
model simulation with all-atom {\it Pulchra}  yields the numerical 
value $\chi^2 = 0.52$. 
At this temperature the mean field model results with higher $k\theta$ {\it i.e.} 
those corresponding to the self-avoiding random walk also produce comparable relatively low $\chi^2$ value,
independently of the value of $k\theta$. The reason for this could be  
in the presence of unusual angles and side-chain orientations in the hydrophobic 
core  of Engrailed ~\cite{Banachewicz-2011}. We may expect that with increased temperature, such unusual orientations
become washed out by thermal fluctuations, making the {\it Pulchra} side-chain reconstruction more reliable. 
We propose this effect is 
observed in Figure \ref{fig-9}, in the $\chi^2$ values at $55 \, ^\mathrm o
\mathrm C$

Indeed, it is not clear how reliable an all-atom library such as {\it Pulchra}, which is designed to model side-chain atoms in
static crystallographic
protein structures at thermal equilibrium,  is in modelling side-chains in a dynamical scenario such as the one considered here.  
Accordingly we can try and eliminate those effects that are concomitant to the {\it Pulchra}  library. For this we
prepare a more rudimental $k\theta = 10^{-9}$ mean field model where we account only for
the C$\alpha$ and C$\beta$  atoms. For this  we simply replace all side-chains along the  Engrailed backbone
by (mock) alanine side-chains using {\it Pulchra}.  
A comparison between this simplified
model and  experimental S/WAXS data, for each of the four temperature values, 
is also shown in the Figure \ref{fig-9}; see the individual starred entries.
For the more rudimental C$\alpha$-C$\beta$  (alanine) mean field model the  $\chi^2$ values are clearly smaller, than
for the all-atom {\it Pulchra} model; the values are very close to $\chi^2 \approx 0.3$ for the temperature range 30 $^{\rm o}$C$-$55 $^{\rm o}$C.  Such a small 
$\chi^2$ value is  {\it very} close to the background noise limit in the experimental data. Thus a deviation
of the simulated data from the experimental data becomes indifferent, at this low level of $\chi^2$ values: At the level of 
C$\alpha$ backbone and including the C$\beta$ side chain atoms only, our simulated results coincide
with experimental data, essentially within the error-bars in the latter.  
Note that when T~=~ 20 $^{\rm o}$C 
we obtain the slightly higher $\chi^2 \approx 0.76$ in the case of  the C$\alpha$-C$\beta$ backbone, but this is also
a clear improvement over the corresponding all-atom {\it Pulchra} value. See Table \ref{table-3} for a summary.

We also estimate the $\chi^2$  values using MD simulations. However, due to limitations in
available computer power a fixed-temperature {\it goodness-of-fit} analysis is not possible. Instead, following ~\cite{Nasedkin-2015} 
we perform an exhaustive fit  using a {\it single} data pool containing {\it all} the 
available MD trajectories between $275$K and $500$K. We then compare this {\it mixed } pool with the individual
experimental data, at the various temperature values. The results are summarised in Table \ref{table-3}. We
note a clear drop in  the values of $\chi^2$ at T~=~55 $^{\rm o}$C, when the structures apparently 
start transiting from the collapsed phase to 
the self-avoiding random walk phase. 
%
%
%
%
%
%
%
%
{
\begin{table}[htb]
       \centering
\begin{tabular}{|c||c|c|c|c|}
\hline  $^\mathrm o$C & 20 & 30 & 40 & 55   \tabularnewline
\hline $\chi^2_{\rm MFM} $ & ~ 0.76 ~ & ~ 0.36 ~ & ~ 0.31 ~ & ~ 0.3 ~ \tabularnewline
\hline $\chi^2_{\rm MD}$  & 0.72 & 0.76 & 0.47 & 0.31 
  \tabularnewline
\hline \end{tabular}
\caption{\small
\it $\chi^2$ values for the rudimental C$\alpha$-C$\beta$ mean field model  (MFM) and MD
model. In the case of the mean field model, we use the results with $k\theta= 10^{-9}$ to compare with experimental
S/WAXS data. In the case of MD, we combine all available data over the entire 
temperature range $T=275-500$K into a single pool of (statistical) data
 }
       \label{table-3}
\end{table}
}
%
%
%
%

In summary, our {\it goodness-of-fit} analysis shows that at the level of the rudimental 
C$\alpha$-C$\beta$ (alanine) backbone, the mean field model simulation 
reproduces the experimental data with very high precision, over an extended temperature range.
The simulated results are essentially within the background noise range of the experimental measurements. 
At the all-atom level the combination of the mean field model with {\it Pulchra} 
results in higher $\chi^2$ value. But we note that  all-atom reconstruction programs such as {\it Pulchra}  are 
primarily intended to model static crystallographic protein structures at very low temperatures, not protein dynamics.

%
%
%
%
%

\subsection{S/WAXS scattering curves}

In Figure \ref{fig-10} (top) we show how the physiologically relevant mean field 
model simulation at $k\theta = 10^{-9} \sim 35 ^\mathrm o 
\mathrm C$ 
in combination with {\it Pulchra} all-atom reconstruction, fits the experimentally measured
S/WAXS scattering data. We note that the overall quality of the mean field model result is fully 
comparable to that obtained with the exhaustive MD {\it mixed} pool 
fitting ~\cite{Nasedkin-2015}, shown in Figure \ref{fig-9} (bottom). 
Besides, the mean field model result is available at individual temperature values and over a wide
range of temperatures, which is difficult to achieve in the case of MD using the presently available computational
resources.  As in the case of Figure \ref{fig-9} we note that the quality of the mean field model result is 
presumably hampered by the need to use an all-atom reconstruction 
procedure such as {\it Pulchra}, which places the side-chain atoms into their optimal crystallographic (low temperature)
thermal equilibrium positions.  

 \vskip 0.3cm
  
\section{Conclusions}
Small to wide angle x-ray scattering (S/WAXS) experiments are emerging as powerful methodology to
analyse protein structure and dynamics. As a consequence there is a need to develop computational tools,  
for structure reconstruction and analysis of S/WAXS data, and for efficient comparison between measurements and
theoretical predictions.  All-atom molecular dynamics remains the 
most comprehensive and reliable method to describe the dynamics of a protein, with atomic level scrutiny. 
However, the enormous demand that it places on of computational power, in particular in a dynamical situation, 
makes it difficult to productively employ 
MD in the interpretation of experimental data and structure reconstruction, in solution X-ray scattering experiments. 
Here we have found 
that a mean field approach in combination with methods of non-equilibrium statistical mechanics, can provide a 
pragmatic and computationally  highly effective, complemental approach to describe and interpret
data  in an S/WAXS experiment. Apparently, the mean field model can reach a very good precision over an extended 
temperature  range, with minimimal need of computational  capacity. We look forward for future S/WAXS
experiments that cover a wider temperature range, to compare with our simulation predictions.

\vskip 0.2cm

\section{Acknowledgements}

This work was supported by
Region Centre Recherche d'Initiative Academique grant, Sino-French
Cai Yuanpei Exchange Program (Partenariat Hubert Curien),
Vetenskapsr\aa det, Carl Trygger's Stiftelse f\"or vetenskaplig forskning, and Qian Ren Grant at BIT.
The computations were performed on resources provided by the Swedish National Infrastructure for Computing (SNIC) at the High Performance Computing Center North (HPC2N).
We thank J. Dai for help in the analysis.

\vskip 0.3cm

\section{Appendix A}

Various elastic network models \cite{Bahar-1997,Halilgolu-1997,Atilgan-2001,Bahar-2010,Lezon-2010,Srivastava-2012,Srivastava-2016} 
have been introduced, to describe aspects of protein dynamics.  They have
met success in modelling small amplitude oscillations around crystallographic protein structures, in particular
the B-factors. Thus it is of interest to
investigate to what extent an elastic network model could be employed to approximate the large scale
structural changes that are observed in S/WAXS experiments. In particular, we are motivated to explore connections
between (\ref{eq:A_energy}) and elastic network models.

%
%
%

We simplify our analysis and use the reality, that over a biologically relevant time scale the
distance between two neighbouring C$\alpha$ atoms has the value $\sim 3.8$ \AA;   the exceptions 
are {\it cis}-peptide planes, which are rare. The entire C$\alpha$ backbone can then be reconstructed in terms of its virtual
bond and torsion angles ($\kappa_i, \tau_i$) that appear  
in  (\ref{eq:A_energy}), using the discrete Frenet equation \cite{Hu-2011}.

We note that the variant of elastic network model considered in 
\cite{Srivastava-2012,Srivastava-2016} employs angular
variables ($r_{ab}, \theta_{abc},\phi_{abcd}$) that relate to  the ($\kappa_i, \tau_i$)  in 
(\ref{eq:A_energy}). Consequently we base our analysis on 
the variables in \cite{Srivastava-2012,Srivastava-2016}; however,  we consider an energy function
that has a more general form. The variables in  \cite{Srivastava-2012,Srivastava-2016} are as follows:

For each C$\alpha$ pair ($a,b$) that are not nearest neighbour along the backbone, the
$r_{ab}$ is their distance. When the structure of peptide planes is fixed, this distance is in effect 
a function of the bond and torsion angles {\it i.e.} we have 
$r_{ab}=r_{ab}(\kappa_i,\tau_i)$ along the C$\alpha$ backbone.

For each triplet \{$a,b,c$\} of C$\alpha$ atoms, the $\theta_{abc}$ is the virtual bond angle; 
note that \{$a,b,c$\}  do not need to be back-to-back along the C$\alpha$ backbone.
These angles are then functions of the C$\alpha$ backbone bond and torsion angles {\it i.e.}
$\theta_{abc} = \theta_{abc}(\kappa_i,\tau_i)$.

For each four consecutive C$\alpha$ atoms $<\!\! a,b,c,d\!\!>$  we have the C$\alpha$ backbone
torsion angles $\phi_{abcd}$. Thus, these coincide with the torsion angles $ \tau_i$
in (\ref{eq:A_energy}).

Note that the set ($r_{ab}, \theta_{abc} , \phi_{abcd}$) is over-complete for the C$\alpha$ backbone,
in general there are many more variables than there are C$\alpha$ coordinates ($\kappa_i,\tau_i$).
For the thermodynamical minimum energy conformation these
variables have the values
\[
\xi_\alpha \sim (r_{ab}, \theta_{abc} , \phi_{abcd}) \ \longrightarrow \ (\hat r_a, \hat\theta_{abc},  \hat\phi_{abcd}) \sim \hat \xi_\alpha
\]
where we introduce $\xi_\alpha$ as a collective of the ($r_{ab}, \theta_{abc} , \phi_{abcd}$).
We are interested in the relevant free energy $F$ around its minimum value.
We denote the ensuing deviations in the variables by
\begin{equation}
\begin{matrix} 
& \delta r_{ab} & = & r_{ab} - \hat r_{ab} &  ~ \\
& \delta \theta_{abc} & = & \theta_{abc} - \hat\theta_{abc} & ~ \\
& \delta \phi_{abcd} & = &  \phi_{abcd} - \hat \phi_{abcd} & \equiv \tau_i - \hat \tau_i
\end{matrix}
\label{diffs}
\end{equation}
Collectively,
\[
\delta \xi_\alpha = \xi_\alpha - \hat\xi_\alpha
\]
The Taylor expansion of the free energy  around the minimum starts with
\[
F[ r_{ab}, \theta_{abc} , \phi_{abcd} ] \equiv F[\xi] \ = \  F[ \hat \xi ] \ +
\]
\begin{equation}
+ \sum\limits_\alpha  \, \frac{\partial F}{\partial \xi_\alpha }_{|\hat\xi}\! \!\! \delta \xi_\alpha + 
\frac{1}{2} \sum\limits_{\alpha\beta} \, \frac{\partial^2 F}{\partial \xi_\alpha \partial \xi_\beta}_{|\hat\xi}\!\!\! \delta \xi_\alpha 
\delta\xi_\beta + \mathcal O (\delta \xi^3)
\label{expa}
\end{equation}
The first term  evaluates the free energy at the minimum. Since $\hat\xi_\alpha$ 
correspond to this minimum the second term should vanish so that we are left with the following leading order correction
to the free energy,
\begin{equation}
\delta F(\xi) \ = \ F(\xi) -  F(\hat\xi) = \frac{1}{2} \sum\limits_{\alpha\beta} \, \frac{\partial^2 F}{\partial \xi_\alpha \partial \xi_\beta}_{|\hat\xi}\!\!\! \delta \xi_\alpha 
\delta\xi_\beta 
\label{Fene}
\end{equation}
The variables $r_{ab}$ and $\theta_{abc}$ have in general a complex dependency on the  
C$\alpha$ coordinates ($\kappa_i,\tau_i$), but implicit in (\ref{Fene}) is the assumption that (at least)  
to the leading order we have
\[ 
\delta r_{ab} \ \approx  \ \sum_i \frac{\partial r_{ab}}{\partial \kappa_i}_{|\hat r} \delta \kappa_i + 
\frac{\partial r_{ab}}{\partial \tau_i}_{|\hat r}  \delta \tau_i \ + \ \mathcal O(\delta \xi^2)
\]
\[ 
\delta\theta_{abc} \ \approx  \ \sum_i \frac{\partial \theta_{abc}}{\partial \kappa_i}_{|\hat \theta} \delta \kappa_i + 
\frac{\partial \theta_{abc}}{\partial \tau_i}_{|\hat \theta}  \delta \tau_i \ + \ \mathcal O(\delta \xi^2)
\]
and 
\[
\delta \phi = \delta \tau
\]
since the torsion angles coincide.
Thus, we find to the leading order the following generic elastic network model expression of the free energy,
\begin{equation}
\delta F(\kappa,\tau) \ = \ \sum_{i,j}\left\{  \Gamma^{\kappa\kappa}_{ij}  \delta\kappa_i \delta\kappa_j + 
\Gamma^{\kappa\tau}_{ij} \delta\kappa_i \delta\tau_j  + \Gamma^{\tau\tau}_{ij} \delta \tau_i \delta \tau_j \right\}
\label{el-net}
\end{equation}
with connectivity matrices $\Gamma^{\kappa\kappa}, \Gamma^{\kappa\tau}, \Gamma^{\tau\tau}$ that are independent of the
backbone coordinates ($\kappa_i,\tau_i$).  Different choices of $\Gamma$ specify different elastic network models.
Commonly,  the connectivity matrix is taken to vanish when the spatial distance $|\mathbf x_i - \mathbf x_j|$
between the C$\alpha$ 
carbons $i$ and $j$ exceeds a prescribed value. 
Note that we may also write (\ref{el-net})  in terms of the variables ($\kappa,\tau$) in (\ref{eq:A_energy}), as follows
\begin{equation}
= \sum_i \{ A_i \kappa_i + B_i \tau_i \} + \sum_{ij} \{ 
C_{ij} \kappa_i \kappa_j + D_{ij} \kappa_i \tau_j + E_{ij} \tau_i \tau_j\} 
\label{exp-F}
\end{equation}
with $\kappa,  \tau$ independent  but in general $|\mathbf x_i - \mathbf x_j|$ dependent
connectivity matrices $A, \dots , E$. 

In the case of (\ref{eq:A_energy}), the functional form of the energy follows from symmetry considerations
~\cite{Chernodub-2010,Molkenthin-2011,Niemi-2003,Danielsson-2010,Hu-2011a,Krokhotin-2012-1,Krokhotin-2012-2,Krokhotin-2013-2,Niemi-2014}. In particular, the principle that the energy should remain 
invariant under local frame rotations is exploited to arrive at the functional form  (\ref{eq:A_energy}).  
We now inquire whether similar symmetry considerations could be introduced to 
deduce a nonlinear extension of the energy function  (\ref{el-net}), (\ref{exp-F}), in some kind of a natural fashion. 
For this  we start with  the following  complex valued quantity
\begin{equation}
\mathcal F(\Psi) \ \sim \ \sum_{ij}  \Psi_i^\dagger  V_{ij}(|\mathbf x_i - \mathbf x_j|) \Psi_j 
\label{PsiV}
\end{equation}
where 
\begin{equation}
\Psi_i  \ \equiv \ \Psi(\mathbf x_i)  
= \left( \begin{matrix} e^{i\varphi_{12}} \cos \vartheta   \\   e^{i\varphi_{34}}  \sin\vartheta \end{matrix} \right)
\label{Psi}
\end{equation}
The free energy of interest is then a linear combination of real and imaginary parts of (\ref{PsiV}).

We note that (\ref{PsiV}) engages a structure akin a O(4) spin glass model, in a spinorial 
representation \cite{Hu-2013,Ioannidou-2014,Ioannidou-2016,Gordeli-2016}. 

For a symmetry principle, we demand that (\ref{PsiV}) 
should remain invariant under a local U(1) rotation that sends
\begin{equation}
\Psi_i \ \equiv \ \Psi (\mathbf x_i)  \ \to \ e^{i\eta(\mathbf x_i) } \Psi(\mathbf x_i)
\label{U1}
\end{equation}
We take the connectivity matrix $V$ in (\ref{PsiV}) to have the form 
\begin{equation}
V(|\mathbf x_i - \mathbf x_j|)  =  
\left( \begin{matrix} \vspace{0.1cm} \rho^{11}_{ij} &  \rho^{12}_{ij}  \\  
 \rho^{21}_{ij} &  \rho^{22}_{ij}  \end{matrix} \right)\cdot
 e^{\ i\!  \int\limits_{\mathbf x_j}^{\mathbf x_i} \mathbf A \cdot d\mathbf x} 
 \label{V-A}
 \end{equation}
with $\rho^{ab}_{ij} = \rho^{ab}(|\mathbf x_i -  \mathbf x_j|)$.
When we choose the vector field $\mathbf A(\mathbf x)$ to transform according to
\begin{equation}
\mathbf A(\mathbf x) \ \longrightarrow \ \mathbf A(\mathbf x) + \nabla \eta
\label{A-gau}
\end{equation}
under the U(1) rotation (\ref{U1}), 
the functional  (\ref{PsiV}) remains intact under the combined transformation (\ref{U1}), (\ref{A-gau}).
For consistency, to ensure that (\ref{V-A}) is independent of the path connecting $\mathbf x_i$ and $\mathbf x_j$,
we demand that \cite{Gordeli-2016}
\[
\nabla \times \mathbf A = 0 \ \ \ \ \Rightarrow \ \mathbf A = \nabla \Phi \ \ \ {\rm (locally)}
\]
We proceed to identify U(1) invariant  combinations of the variables in (\ref{Psi}):  
With ($\sigma_1,\sigma_2,\sigma_3$) the standard Pauli matrices we define the three component unit vector
\begin{equation}
\hat{ \mathbf n } \ = \ \Psi^\dagger \hat \sigma \Psi \ \Rightarrow \ \mathbf n = \left( \begin{matrix} \sin 2\vartheta 
\cos (\varphi_{34}-\varphi_{12}) 
\\  \sin 2\vartheta  \sin (\varphi_{34}-\varphi_{12}) 
\\ \cos 2\vartheta \end{matrix} \right)
\label{unit-n}
\end{equation}
Clearly, this vector is invariant under the U(1). 

We note that the torsion angle $\tau$ in (\ref{eq:A_energy}) has a natural interpretation as the longitude on a two-sphere, 
and the bond angle $\kappa$ has similarly a natural identification as the latitude
~\cite{Chernodub-2010,Molkenthin-2011,Niemi-2003,Danielsson-2010,Hu-2011a,Krokhotin-2012-1,Krokhotin-2012-2,Krokhotin-2013-2,Niemi-2014}. Thus these variables can be combined into a three component unit vector such as (\ref{unit-n}),
in a canonical fashion.
 Accordingly we have the U(1) invariant identifications 
\[
\begin{matrix} \tau & = & \varphi_{34} - \varphi_{12} \\
\kappa & = & \hspace{-1.1cm} 2 \vartheta  \end{matrix}
\]
The U(1) gauge transformation (\ref{U1}) then  corresponds to a frame rotation, around the direction of $\mathbf n$
~\cite{Chernodub-2010,Molkenthin-2011,Niemi-2003,Danielsson-2010,Hu-2011a,Krokhotin-2012-1,Krokhotin-2012-2,Krokhotin-2013-2,Niemi-2014}.

We choose the gauge so that
\[
\mathbf A  = \nabla \{  \frac{1}{2} (\varphi_{34} + \varphi_{12}) \}
\]
and we substitute this in (\ref{PsiV}). We work out the real 
and imaginary part of (\ref{PsiV}) separately, and Taylor expand
the trigonometric functions to second order to find
\[
{\rm Re} \, \mathcal F[\kappa,\tau] \ = \ \frac{1}{2}\sum\limits_{i,j}  \left\{ \rho^{12}_{ij} +
 \rho^{21}_{ij} \right\} \kappa_j 
\]
 \begin{equation}
 - \frac{1}{4} \sum\limits_{i,j}   \rho^{11}_{ij} \left\{ (\tau_i - \tau_j)^2 + \kappa_j^2 \right\}
+ \frac{1}{4} \sum\limits_{i,j}   \rho^{22}_{ij}   \kappa_i \kappa_j 
 \label{reF}
 \end{equation}
\begin{equation} 
%
{\rm Im} \ \mathcal F[\kappa,\tau] = \frac{1}{2} \sum\limits_{i,j}  \left\{ \rho^{12}_{ij} -
 \rho^{21}_{ij} \right\} (\tau_i + \tau_j) \kappa_j 
 \label{imF}
 \end{equation}
A frame independent energy function akin  (\ref{exp-F}) 
can  then introduced, as a linear combination of (\ref{reF}), (\ref{imF}) and 
with $\rho^{ab}_{ij}$ the connectivity matrix.

There is a notable conceptual differences between (\ref{eq:A_energy}), and (\ref{exp-F}): 
The free energy (\ref{exp-F}) is a quadratic function of the variables while (\ref{eq:A_energy}) is 
quartic. Thus the soliton that constitutes the hallmark of the DNLS equation and models a protein 
loop in the approach based on (\ref{eq:A_energy}), is absent  in conventional (linear) elastic network models;
the loops have a different origin, in the latter. 

Moreover,  unlike (\ref{eq:A_energy}) where only  
nearest neighbour couplings appear explicitely, in the elastic network models there can be a
direct, space coordinate dependent coupling between {\it any}  two amino acids $i$ and $j$.
As a consequence,  in an elastic network model the number of parameters is
commonly much larger than the number of independent C$\alpha$
coordinates. For example \cite{Srivastava-2016} estimates in concrete examples,
that the number of  independent angular variables 
in their elastic network model is about 15-17 times larger than the number of independent backbone angles. 
On the other hand, in the case
of (\ref{eq:A_energy}), the number of parameters is commonly comparable to the number of amino acids which
yields much tighter experimental constraints on the model.

The commonly employed  linear elastic network models are by their 
design describing the small amplitude 
motions around a protein conformation that corresponds to a minimum 
of free energy, such as a crystallographic protein structure. 
These models are not even 
designed to describe the large conformational deformations that one encounters  
in S/WAXS experiments, in their present form 
they are not intended to describe large structural deviations from the
minimum energy \cite{Bahar-2010,Lezon-2010}. 

We conclude that elastic network models have their conceptual foundation in spin 
glass models \cite{Bryngelson-1987} while
the energy function (\ref{eq:A_energy})  builds on the concept of collective oscillations and large scale 
structure formation in conventional nonlinear dynamics.
Accordingly, these two approaches are  complementary to each other. 
In a future publication we shall  investigate the
O(4) spin glass model (\ref{PsiV}), to systematically 
account for non-linear corrections in the context of the elastic network model 
along the lines of the DNLS model.

\end{document}